\documentclass [12pt]{revtex4}
\usepackage{amssymb}

\usepackage{amsmath}
\usepackage{amsfonts}

\begin{document}
\title{A Pedestrian Approach to the Measurement Problem in Quantum Mechanics}

\author{Stephen Boughn}

\affiliation{Departments of Physics and Astronomy, Haverford College, Haverford, PA 19041}

\author{Marcel Reginatto}
\address{Physikalisch-Technische Bundesanstalt, Braunschweig, Germany}

\date{\today}

\begin{abstract}

The quantum theory of measurement has been a matter of debate for
over eighty years. Most of the discussion has focused on theoretical
issues with the consequence that other aspects (such as the
operational prescriptions that are an integral part of experimental
physics) have been largely ignored. This has undoubtedly exacerbated
attempts to find a solution to the ``measurement problem". How the
measurement problem is defined depends to some extent on how the theoretical
concepts introduced by the theory are interpreted. In this paper, we
fully embrace the minimalist statistical (ensemble) interpretation
of quantum mechanics espoused by Einstein, Ballentine, and others.
According to this interpretation, the quantum state description
applies only to a statistical ensemble of similarly prepared systems
rather than representing an individual system. Thus, the statistical
interpretation obviates the need to entertain reduction of the state
vector, one of the primary dilemmas of the measurement problem. The
other major aspect of the measurement problem, the necessity of
describing measurements in terms of classical concepts that lay
outside of quantum theory, remains. A consistent formalism
for interacting quantum and classical systems, like the
one based on ensembles on configuration space that we
refer to in this paper, might seem to eliminate this facet of the
measurement problem; however, we argue that the ultimate interface
with experiments is described by operational prescriptions and not in terms of the concepts of classical theory. There is no doubt that attempts to address the measurement problem have
yielded important advances in fundamental physics;
however, it is also very clear that the measurement problem
is still far from being resolved.  The pedestrian approach presented here
suggests that this state of affairs is in part the result of searching for a
theoretical/mathematical solution to what is fundamentally an
experimental/observational question.  It suggests also that the measurement problem is, in some sense, ill-posed and
might never be resolved.  This point of view is tenable so long as one
is willing to view physical theories as providing models of nature
rather than complete descriptions of reality. Among other things,
these considerations lead us to suggest that the Copenhagen
interpretation's insistence on the classicality of the measurement
apparatus should be replaced by the requirement that a measurement,
which is specified operationally, should simply be of sufficient
precision.
\end{abstract}

\maketitle

\section{Introduction}\label{INTRO}

Since the beginning of quantum mechanics more than 80 years ago,
physicists have argued about how to interpret the theoretical
concepts introduced by the theory. Perhaps the most troublesome of
all is the meaning of the wave function $\Psi$
 introduced by Schr\"{o}dinger in 1926 \cite{S2003}. Both the
statistical rule, probability $\sim  |\Psi|^{2}$, suggested by Born
\cite{B1926} and the concomitant phenomenon of quantum interference were
anathemas to classical physics. The founders of quantum mechanics
including Bohr, Heisenberg, Pauli, Schr\"{o}dinger, and
Einstein spent considerable effort worrying about how best to
interpret the theory. In fact, there is more than one
``interpretation'' of quantum mechanics and the precise meanings of
these remain the subject of much discussion. There has never been
(nor, perhaps, ever will be) complete agreement on this issue.

Bohr's point of view, commonly referred to as the {Copenhagen
interpretation} (see below), was for many years considered to be the agreed
upon viewpoint, at least as was declared in standard textbooks on quantum
mechanics. Modern texts and the teachers that use them are often more
circumspect and usually list a variety of interpretations such as the ensemble interpretation, the Copenhagen interpretation, decoherence
theory, realist models such as Bohmian mechanics, the many worlds
interpretation, etc. The truth of the matter is that few physicists actually
know the details of any of these interpretations.  Even the principal
proponents of the Copenhagen interpretation, Bohr, Heisenberg, and Pauli
disagreed on various aspects of it. Most physicists seem to arrive at
some vague personal interpretation of quantum mechanics and then stop
worrying about it, following the David Mermin maxim embodied in his
statement, ``If I were forced to sum up in one sentence what the
Copenhagen interpretation says to me, it would be {`Shut up and
calculate!'}'' \cite{M1989}. That is, if one solves
Schr\"{o}dinger's equation (or a relativistic equivalent) and uses
the statistical Born rule to interpret the solution, then quantum
theory seems to provide a complete description of all that can be
observed. Further discussion as to the physical significance of
$\Psi$ or as to whether or not it provides a complete description of
reality is unnecessary and should be eschewed.

As a case in point, consider the so called Copenhagen interpretation.
While physicists still argue about what is and what is not included
in the Copenhagen interpretation, at the very least most would agree
that, according to this interpretation, the wave function, or
alternatively the density matrix, predicts the probability
distribution of outcomes of particular measurements made on an
ensemble of similarly prepared systems. However, as to how to
perform these measurements, both theory and the accompanying
interpretation are silent. Such predictions must be turned over to
experimental physicists in whom both classical physics and the art
of constructing apparatus are deeply ingrained. (While not
usually acknowledged, the same is true in other domains of physics.
That is, even after an interpretation is given, there is no
implied prescription for how to perform an experiment.)  Nevertheless,
the Copenhagen interpretation often serves as a foil
for discussions of interpretations of quantum mechanics and the
measurement problem and it is useful to make explicit those aspects
of it that are referred to in this paper, with the caveat that there
is no well defined Copenhagen interpretation. In Heisenberg's words:
``...it may be a point in the Copenhagen interpretation that its
language has a certain degree of vagueness, and I doubt whether it
can become clearer by trying to avoid this vagueness.'' \cite{S1972}
As mentioned above, the main proponents didn't fully agree on
what the Copenhagen interpretation entails. So we use the term
simply to label the following interpretative statements that are
often attributed to it and other interpretations, and to which
many physicists ascribe.

According to Stapp \cite{S1972}, ``The logical essence of the Copenhagen
interpretation is summed up in the following two assertions: 1) the
quantum theoretical formalism is to be interpreted
\textit{pragmatically}; and 2) quantum theory provides for a
\textit{complete} scientific account of atomic phenomena.'' Several
operational principles that are often associated with the Copenhagen
interpretation are: i) the square of the magnitude of the wave
function, $\vert \Psi \vert ^{2}$, is associated with the
probability of the occurrence of an event (the Born rule); ii) it is
not possible to determine, via measurement, all of the possible
properties of a given system (the Heisenberg uncertainty principle); iii)
the only possible values of a given property that can result from a
measurement are the eigenvalues of the operator associated with that
property; and iv) measurements must be (or invariably are) made
with apparatus that are described in terms of classical physics.
Item iii) invariably leads to wave function collapse (see below);
however, this is not problematic for the Copenhagen or ensemble
interpretations for which the wave function is considered
to be a computation tool rather than an aspect of physical reality.
This list is by no means exhaustive but will suffice for the
purposes of this paper.

Perhaps the single most perplexing aspect of interpreting quantum
mechanics is what is generally referred to as the
\textit{measurement problem}, i.e., the unresolved problem of how
the outcome of a particular measurement arises from a quantum theory
that, at most, renders a probabilistic distribution of all possible
outcomes. We say ``at most'' because quantum theory itself says
nothing at all about the measurement process. The probabilistic
significance of the result of a quantum mechanical calculation
arises from interpretive statements that accompany quantum mechanics
but such statements are not, in themselves, intrinsic to the theory.
For example, most interpretations maintain that the
measurement of any observable of a system can only be an eigenvalue
of the Hermitian operator associated with that variable and the
probability of measuring a particular value is given by the absolute
value squared of the corresponding eigenfunction,
$|\lambda_i\rangle$, projected onto the wave function,
$|\Psi\rangle$, of the system, i.e.,
$|\langle\lambda_i|\Psi\rangle|^2$, the {\it Born rule}. However,
such an interpretation is not intrinsic to the mathematical
structure of the theory.  Furthermore, just what constitutes a
measurement of an observable is not well defined and, in fact,
constitutes one of the aspects of the measurement problem. We
maintain that this is also the case for classical theory and contributes
to a ``classical measurement problem'' (see Section \ref{MPCP}).  There
are other interpretations that offer different explanations as to
the significance of the wave function and they are all extrinsic to
the mathematical formalism of quantum mechanics and are all vague
about the details of what constitutes a measurement. Different
people have emphasized different aspects of the measurement problem;
however, the following four (related) issues are frequently raised:

1) \textit{Wave Function Collapse}: Part of any standard
interpretation of quantum mechanics is that the only possible
outcomes of the measurement of a particular property of a system are
eigenvalues of the quantum mechanical operator associated with that
property. Even though it is not part of the Copenhagen
interpretation, many physicists harbor the belief that the wave
function represents the real world and, furthermore, that it
provides a complete description of an individual system. If this
were so, then after a measurement has been made, it must be that the
wave function of a system is transformed by the measurement from the
initial wave function to the eigenfunction associated with the
measured value. The problem is that such a transition is not part of
the unitary evolution of the system as described by quantum theory.
If it were, then presumably quantum theory would predict this
transition and, hence, the exact outcome of the measurement, thereby
contradicting the statistical interpretation that lies at the heart
of the theory. This aspect of the measurement problem is confounded
by the fact that wave function collapse, if assumed to be a physical
process, would not be expected to occur instantaneously and should
therefore be accessible to observation. Decoherence theory has been
useful in understanding how the wave function of a system evolves as
it interacts with a measuring device and the environment if these are also
described quantum mechanically by a wave function.  However, this
evolution is necessarily distinct from wave function collapse \cite{S2004}.

2) \textit{The Consistency of Quantum and Classical Reality}: In
some sense, this is the flip side of the first issue.  We perceive
reality as a series of events involving the objects of our
perceptions.  These occur sequentially and definitively.  On the
other hand, quantum mechanics seems to say nothing what-so-ever
about these events but only describes, exactly, the evolution of
interacting wave functions.  An interpretation of the wave functions
is necessary to make the connection with events and, in the end, it
is only a probabilistic statement about the {\it real} world.  In
short, the formalism of quantum mechanics precludes the occurrence
of any specific event whereas, in our world, we know that specific
outcomes always occur.

3) \textit{Classicality of Experiments}: According to Bohr's
statement of the Copenhagen interpretation, while a (microscopic)
system under investigation is described quantum mechanically, the
measurement apparatus that observes it must be (or, perhaps, always
is) described classically. How does one decide which aspects of a
system are to be described classically and which to be described
quantum mechanically, i.e., what is the location of the
quantum/classical divide? Many, including Heisenberg, have
pointed out that this divide does not represent a discontinuity
of 	physical systems but rather is simply a transition from one
formalism to another.	Nevertheless, it has historically been
considered to be an important aspect of the measurement problem.

4) \textit{Interference Effects}:  One of the most amazing
consequences of the quantum nature of matter is quantum
interference, a phenomenon in which a particle's wave function
exhibits wave interference.  For example, if a single particle wave
function passes through a barrier with two slits, the parts of the
wave function emerging from the two slits interfere and there will
be periodic locations on a distant screen where there is (near) zero
probability that the particle will strike. If one of the slits is
covered, the interference fringes disappear. One wonders why, by
closing off one possible path of the particle, the probability of
it striking the screen at a previously inaccessible region becomes
nonzero? This sort of interference is not the
least bit surprising for inherently wave phenomena like sound or
light; however, that ``particles'' should behave this way runs
counter to our intuition. Furthermore, if one simply determines
through which slit the particle passes without obstructing it, then
quantum interference between the two parts of the wave function
disappears. The only way this can be explained is that such a
determination, in some sense, constitutes a measurement and the wave
function of the particle collapses.

One rather curious aspect of the measurement problem should be kept
in mind. It can certainly be argued that for the last 80 years,
while many interesting resolutions of the problem have been
suggested, the problem of quantum measurements remains largely
unsolved. Yet, the advance of quantum mechanics and quantum field
theory has been enormous and seems not to have been impeded in the
least by the lack of a satisfactory resolution of the measurement
problem nor even by the lack of an agreed upon interpretation of the
theory. How can this be? On the other hand, we will show that this
situation is not particularly remarkable in the context of the
pedestrian approach presented here.

\section{Decoherence Theory}

Today, many physicists are of the mind that decoherence theory
has largely resolved the measurement problem. While certainly
relevant to the measurement problem, it is a far reach indeed to
claim that decoherence theory has solved the problem. Decoherence
theory is relevant to those aspects of the measurement problem that
deal with the classicality of macroscopic measurement apparatus and
it will be important to compare the perspective of decoherence
theory with that of the pedestrian approach put forward in this
paper. The details of decoherence theory would be much too large a
diversion to undertake in this paper; however, there are many
accessible treatments in the literature including the books by Joos
et al \cite{J2003} and Schlosshauer \cite{S2007}.

Decoherence theory is neither new physics nor a new interpretation
of quantum mechanics; although, it is certainly relevant to
questions of interpretation. In decoherence theory, the measuring
apparatus and the environment with which it inevitably interacts are
both treated as purely quantum mechanical systems. As a consequence
of the interactions of the quantum system of interest with the
measuring apparatus and it with its immediate environment, the three
become {entangled}, i.e., strongly correlated with each
other. All, or at least most, of the environmental quantum degrees
of freedom are not observable (certainly, not observed) and,
therefore, must be summed over to achieve a reduced state of the
system plus apparatus. The net effect of the enormous number of
environmental degrees of freedom is that off-diagonal terms of the
reduced density matrix rapidly vanish, i.e., coherence between the
different eigenstates of the system/apparatus is lost.
Thus, decoherence theory demonstrates why it is that quantum
coherence is seldom,
if ever, observed at the classical (macroscopic) level, the fourth
aspect of the measurement problem listed above \cite{S2004}.

Another aspect of the measurement problem that we have referred
to is the
Copenhagen interpretation's requirement of a classical measurement,
raising the immediate question as to what determines the divide
between the quantum system and the classical measuring apparatus. It
was Heisenberg's view \cite{SC2008} that the
dividing line between quantum and classical did not signify a
discontinuity of the physical process but rather is defined by the
nature of the measurement, which to a certain extent is at the
discretion of the observer.  In fact, it’s not that the measuring 
apparatus is ``classical''
but rather that classical physics formalism is used to analyze its
behavior.  The problem is how to merge the formalisms of quantum
and classical physics so that the evolution of the systems can be
followed through the measurement process.  The Copenhagen and
orthodox von Neumann interpretations lack a description of the
interaction of systems across the quantum/classical divide and
are, therefore, of no help in this respect.  Decoherence theory has, 
in a sense, resolved the dilemma but only by
treating the macroscopic measuring apparatus as a quantum mechanical
system that interacts with the original quantum system via a quantum
mechanical Hamiltonian.  However, one is left with the problem of
the interpretation of the measuring apparatus wave function in terms of
ordinary experience. One still must apply the Born rule of
the Copenhagen interpretation, which, in effect, assumes that only
one of the outcomes actually occurs. In this sense, decoherence
theory does not address the problem that quantum mechanics alone is
insufficient to explain why we do not experience mixed states in our
classical world \cite{S2004}.

Finally, while decoherence theory is relevant to
the measurement process in general and is a useful
computational tool for characterizing many microscopic and
mesoscopic systems of interest (e.g., in foundations of physics,
quantum information, and quantum
computing), it is far from useful in designing {\it most} experiments.
Experimentalists have quite successfully created experiments with no
consideration whatsoever of the form of the apparatus/environment
interaction Hamiltonian. Indeed, in many cases the experiment is
designed so that the coupling of the apparatus and measured system
is strong enough to be able to {neglect} the effects of the
environment. This is highly desirable; in an ideal experiment, the
environment would play no role. Of course, one might say that
experimentalists have simply {learned} by trial and error how to
design experiments, after which these skills simply become part of
one's physical intuition, just as people (even professional
cyclists) learn how to ride bicycles without any knowledge of
rotational dynamics or conservation of angular momentum. These same
comments also apply to the quantum/classical model of interactions
discussed in the Appendix.

\section{System Preparation and Measurement Execution}\label{SPME}

Perhaps because of the emphasis of the Copenhagen interpretation on
measurements, another aspect of the quantum/classical divide is
frequently glossed over and that is {system preparation}.
Preparation and measurement are fundamentally different
\cite{PB1992}. A measurement yields a numerical datum. The repeatable
preparation process generates a statistical ensemble from which data are
collected. But how is it that a complicated and entangled arrangement
of macroscopic classical equipment manages to create a well defined
quantum system that is to be the subject of a subsequent
measurement? A quick check of the two decoherence review articles by
Zurek \cite{Z2003} and Schlosshauer \cite{S2004} reveals no discussion of system
preparation whatsoever. To be fair, the authors do refer to
measurements that leave a system in a specific eigenstate, which can
be viewed as preparing such a quantum system for further
observation. However, in general, system preparation need not
involve a measurement \cite{PB1992} and, indeed, experimentalists probably
wouldn't look at system preparation in this way.

Consider, for example, the prescriptions for preparing an electron
beam for its subsequent use in a double slit interference experiment.
These might include boiling off electrons from a hot filament, which is
heated by current from a power supply, accelerating these
electrons through a known potential (generated from a high voltage generator),
and then passing them through a small aperture so as to approximate a point
source. At large distances from the aperture the electron beam is,
to a good approximation, described as a plane wave momentum
eigenstate. Nowhere in this description were
we required to discuss the quantum nature of electrons. An
apparently classical apparatus was used to
generate a coherent quantum plane wave. One might
consider the electrons initially to be in a pure or even mixed,
state that, by the correspondence principle, only appears to be in a
classical state.  Then system preparation is effected by a
``measuring'' apparatus (e.g., a magnetic field) that is used to select
those electrons that are in the desired
quantum state. However, this view seems to imply that an electron
wave function possesses a reality prior to the experiment and
independent of the experimenter.  If so, then one is led to the view
that wave functions are the real entities which inhabit the universe
with all the concomitant problems (e.g., wave function collapse)
that this entails.

The Copenhagen interpretation does not provide a
straightforward account of system preparation.
In fact, in neither the case of system preparation nor measurement,
are there precise rules for associating the experimental
specifications with the wave functions describing the systems.
In Stapp's ``practical account of quantum theory'' \cite{S1972}, he
emphasizes the operational descriptions of both preparing and
measuring devices. While certainly informed by classical and quantum
theory, the effective ``rules'' of system preparation are arrived at
by calibrating both the system preparation and measuring devices.  The
calibration procedure is facilitated by the leverage of many possible
measurements of the different states of systems.  This leverage was
illustrated by \cite{S1972} with the following example.  Consider
the matrix element between two different systems, $A$
and $B$, with $N_A$ possible eigenstates for the former and $N_B$ for
the latter. Then there are $N_A+ N_B$ unknown functions, $\Psi_A$ and
$\Psi_B$ but $N_A \times N_B$ experimentally determinable quantities,
$| \langle A|B \rangle |^2$.
\begin{quote}
``Using this leverage, together with plausible assumptions about
smoothness, it is possible  to build up a catalog of correspondences
between what experimental physicists do and see, and the wave
functions of the prepared and measured systems.  It is this body of
accumulated empirical knowledge that bridges the gap between the
operational specifications $A$ and $B$ and their mathematical images
$\Psi_A$ and $\Psi_B$.''
\end{quote}

How scientists arrive at these operational prescriptions is an
extremely interesting question that involves theoretical models,
physical intuition, historical precedent and, we suppose, even
sociology and psychology. Is such a topic ever the subject of study?
One might argue that every scientific book and paper ever written
is, in part, an attempt to answer this question.

After a system has been prepared, how does an experimentalist
execute a measurement? In general, the construction of a measurement
apparatus and the subsequent measurement are effected according to
prescriptions that the experimenters have both created themselves
and acquired from others. While the designs of experiments might
well rely on fundamental quantum mechanical and classical
calculations, they also rely on previous observations of the
behavior of systems that are then described phenomenologically,
i.e., not derived from fundamental theory, and in some cases on
conventional wisdom even if that wisdom is not completely
understood. Even then, the ultimate realization of the experiment is
generated from a set of operational prescriptions and the skill of
the experimentalist in realizing the experiment. This is the art of
experimentation. The behavior of all of the equipment generated via
these prescriptions can, in principle, be described by fundamental
quantum (or classical) theory; however, it is doubtful this has ever
occurred when constructing, for example, a soldering iron. Yet, even if
a soldering iron were, somehow, the subject of an experimental
investigation to confirm a quantum mechanical prediction, the test
would presumably require yet another apparatus to perform the
experiment. Our claim is that the prescriptions that define an
experiment are expressed neither in the language of quantum theory
nor in the language of classical theory but rather in the common
(technical) language that directs the actions of the experimenter. 

We should emphasize that these operational prescriptions have little to do
with the stylized experiments and simplified procedures that one might
find in articles on quantum measurement theory.  The latter are those
that can, and invariably are, characterized by the mathematical formalism
of quantum mechanics. (There are good accounts in the literature which
emphasize these more formal aspects of system preparation \cite{L1969, B1998}.)
By operational procedures, we are referring instead to the much more
complicated prescriptions actually used by experimental physicists, engineers,
and technicians when they carry
out their work. While the need for such prescriptions is sometimes addressed
in the literature \cite{S1972, PB1992, P1995}, the prescriptions themselves have
 received surprisingly little attention and the important role that they play is
usually not acknowledged. The expression of operational prescriptions is not
{theoretical} in the sense that there is no well defined set
of consistent mathematical relations that define them. The fact that
nearly all analyzes of the measurement problem avoid discussions of
this type of operational prescription has both been responsible for
the problem being posed in purely ``theoretical'' terms and has
exacerbated attempts to find a solution. This is not a criticism leveled 
at theoretical physicists.
Experimentalists often use similar simplifications in their research
papers.  They frequently give detailed recipes to coworkers or
share crucial prescriptive aspects of their work at meetings but, more often
than not, such details are not discussed in the literature.

The Copenhagen interpretation
seems to tacitly acknowledge this issue by insisting on the classicality of
measurement apparatus, albeit with the ``certain degree of
vagueness'' to which Heisenberg referred \cite{S1972}.
One of the essential elements of the Copenhagen and von Neumann interpretations
is the existence of the quantum/classical divide, the \textit{Heisenberg cut}
if you will.  This postulate asserts that the quantum system is separated from
the classical measuring apparatus by the Heisenberg cut and somewhere on the far
side of this cut there is a classical apparatus and a classical description.
In the Copenhagen/von Neumann interpretation, the measurement problem is sidestepped 
by postulating a correspondence between the quantum world and
the classical description. However, this correspondence is not easy to characterize.
The Copenhagen interpretation lacks a theory that describes the interaction across the cut
and, in fact, presumes that what happens at the cut is not mathematically describable.
Any attempt to describe that interaction in a mathematically consistent way, inevitably
leads to a corresponding measurement theory for classical mechanics which picks up
some of the features of the quantum world. Again, the Heisenberg cut
does not refer to a discontinuity of the physical process but rather to
a discontinuity in the formalism used to treat the system and the measuring
apparatus and the Copenhagen interpretation provides no theory of how to
bridge this divide. 

In the Appendix, we describe an approach that provides a way to merge
the formalisms of quantum and classical theory; however, we maintain that this resolution
does not solve the measurement problem.  There is, in fact, another divide and
that is between quantum/classical formalism and the operational  prescriptions
of experiments.  In some ways, this divide is more insidious in that it is difficult
to imagine a “theory” that would connect the mathematical formalism of theoretical
physics (classical or quantum) with the non-mathematical prescriptions that
define experiments.  While the pedestrian approach of this paper falls short of
resolving the measurement problem (more on this later), it would not make the theory
any less testable nor any less useful than orthodox quantum theory, with its Heisenberg
cut.

An interesting question is why Bohr insisted that
measurements always be described via classical physics. Perhaps it
is not because the measurement process is well described by
classical physical theory, but rather because of the happenstance
that much of the same language is used both for the descriptions of
measurements and the formulation of classical theory.  This might
lead one to the conclusion that all aspects of measurements are well
understood in terms of fundamental classical theory whereas, in fact,
much of our understanding of measuring apparatus is phenomenological
and is supplemented by precisely the sort of operational prescriptions
referred to above.

Before discussing further the role of operational prescriptions and
their impact on the measurement problem, we turn to some aspects of
quantum and classical mechanics that are particularly relevant to our paper. 

\section{Probability and quantum mechanics}\label{PAQM}

Max Born, in the 1926 paper in which he introduced the concept of the probability of a state \cite{B1926a}, gives the following description of the new physics: ``The motion of particles follows probability laws, the probability itself however propagates according to the law of causality.'' In this sentence, as Abraham Pais remarked, Born ``expressed beautifully the essence of wave mechanics'' \cite{P1986}. The probabilistic interpretation developed rapidly; the early history of probability in quantum mechanics may be reconstructed from a footnote in Heisenberg's 1927 paper on the uncertainty relation \cite{H1927} where he lists the main contributions up to that date, starting with Einstein's statistical interpretation of de Broglie waves in his 1925 paper on the quantum gas and continuing with papers of Born, Heisenberg, Jordan, Pauli and Dirac. The literature on this topic is enormous and we will limit ourselves to some historical considerations that are particularly relevant to this paper.

The Born rule sets $P=|\Psi|^2$. It is straightforward to reformulate quantum mechanics so that the probability $P$ plays a central role. As is well known, the polar decomposition $\Psi=\sqrt{P}~e^{iS/\hbar}$ maps the complex Schr\"{o}dinger equation for $\Psi$ to a pair of real, nonlinear equations for two real variables $P$ and $S$ which now become the {\it fundamental variables} of the theory. This transformation was first carried out by Madelung \cite{M1927}. However, as is clear from the title of his paper, his aim was to reformulate quantum mechanics as a hydrodynamic theory and, in accordance with this, he interpreted $P$ as a mass density function, following the first interpretation proposed by Schr\"{o}dinger rather than the new statistical interpretation of Born (Madelung's paper was published in 1927 but submitted in October 1926, a few months after the papers in which Born introduced his new rule). In this formulation, the function $S$ plays the role of a potential for the velocity field $\textbf{v}$ of the fluid, which is defined according to $\textbf{v}=\nabla S/m$. Making use of this interpretation of $S$, Madelung shows that the real and imaginary parts of the Schr\"{o}dinger equation correspond to a hydrodynamical equation of continuity for the mass density function and to an equation for irrotational fluid flow but with a (nonclassical) term which Madelung describes somewhat vaguely as due to the action of ``internal'' forces of the continuum.

Madelung's formulation was revived in the 1950's by Takabayasi \cite{T1952, T1953} and, in modified form, by Bohm and Vigier \cite{BV1954} and Sch\"{o}nberg \cite{S1954}. Takabayasi in particular developed and expanded the theory considerably, presenting it as a new ``hydrodynamical'' quantization procedure. He recognized that full equivalence with wave mechanics required a topological condition which takes the place of single-valuedness of the wave function (i.e., that $\oint_C dS/h$ must be an integer for all loops $C$ in configuration space), introduced the concept of ``quantum stress,'' and extended the theory to spin degrees of freedom, relativistic equations, and fields. Bohm and Vigier further developed the hydrodynamical approach by adding fluctuations to the Madelung fluid and modeling particles as highly localized inhomogeneities that moved with the local fluid velocity. Their physical model, which was presented as an extension of the formulation of quantum mechanics developed by Bohm \cite{B1952}, was in part motivated by criticisms (by Pauli \cite{P1953} and Keller \cite{K1953}) regarding the assumption that the probability distribution of an ensemble of particles coincides with $|\Psi|^2$. The work of Sch\"{o}nberg, while related to the physical picture of the hydrodynamic model, goes far beyond it by introducing the second quantization of the Madelung fluid.

The authors who revisited Madelung's approach in the 1950s considered the physical picture of a fluid to be a useful model. At the same time, they were aware that it was necessary to introduce substantial modifications or reformulations of the original Madelung interpretation to bring it in line with Born's statistical interpretation. That this was Sch\"{o}nberg's motivation for introducing second quantization becomes clear from the following sentence from the introduction of his paper: ``It is well known that the Madelung model does not lead to a satisfactory interpretation of the Schr\"{o}dinger equation. By considering the Madelung formalism as the classical theory of the motion of a fluid medium and applying to it the second quantization we can get a satisfactory interpretation.''

It should also be mentioned that stochastic mechanics, which was introduced by F\'{e}nyes \cite{F1952} in the 1950s and later formulated in a different manner by Nelson \cite{N1966}, may also be seen as a later extension of Madelung's approach, in the sense that the stochastic process that provides the basis of the theory leads to the Madelung equations. This has been emphasized by Guerra, who writes in his review article \cite{G1981} that ``a natural and straightforward particle interpretation of the Madelung fluid is indeed possible, but only by allowing a random character to the underlaying trajectories. In the semiclassical limit $h \rightarrow 0$ the randomness disappears and the trajectories become those of the classical theory, while the Madelung fluid, through the vanishing of the quantum potential, reduces to the Hamilton-Jacobi fluid.''

The most familiar formulation of quantum mechanics which makes use of $P$ and $S$ variables is the de Broglie-Bohm theory \cite{B1952,dB1926,dB1927,dB1927a}. It is, however, conceptually very different from Madelung's formulation and therefore it should not be considered an extension of it. The approach is based on the co-existance of particles and a wavefunction that is assumed to evolve according to the Schr\"{o}dinger equation: instead of replacing the wavefunction by variables $P$ and $S$, the standard quantum mechanical description is {\it completed} by adding point particles which follow definite trajectories which are determined by the wave function. The ontology therefore includes not only particles but also the wavefunction. The field $\textbf{v}=\nabla S/m$ acquires a new interpretation, in that it describes the motion of individual particles rather than the average motion of particles, as in the hydrodynamical approach.

One may adopt a ``minimalist'' approach and drop Madelung's hydrodynamical picture altogether (i.e., the assumption that $P$ is associated with the mass density function of a fluid) and interpret $P$ instead in Bornian fashion as the probability density of particles. We will follow this route here. This shift in interpretation leads to a formulation in which $P$ and $S$ are still fundamental variables of the theory, but now quantum mechanics no longer appears in the guise of a hydrodynamic theory -- it becomes a {\it statistical theory of ensembles on configuration space}. We now give an overview of some of the general features of such theories.

The description of physical systems by ensembles on configuration space (ECS) may be introduced at a very fundamental level, without making reference to quantum mechanics. The starting point is simply a probability density $P(x)$ on the configuration space with coordinates $x$, with $P(x) \geq 0$ and $\int dx \,P(x) = 1$.

To set the probabilities in motion, assume that the dynamics of $P$ are generated by an action principle. This is a rather mild assumption which is valid for a very large class of systems; in particular, it holds for the types of systems that we consider in this paper. We develop the theory using a Hamiltonian formalism for fields (which, for our purposes, is more convenient than using a Lagrangian formalism). Then, to get equations of motion for $P$, introduce an auxiliary field $S$ which is canonically conjugate to $P$ and a corresponding Poisson bracket for any two functionals $F[P,S]$ and $G[P,S]$,
\begin{equation}\label{PoBr}
\left\{ F,G\right\} =\int dx \left\{
\frac{\delta F}{\delta P} \frac{\delta G}{\delta S}
 - \frac{\delta F}{\delta S}  \frac{\delta G}{\delta P} \right\}.
\end{equation}
The equations of motion for $P$ and $S$ take the familiar form
\begin{equation} \label{EqMoPoBr}
\dot{P} = \left\{ P,\tilde{H}\right\} =\frac{\delta
\tilde{H}}{\delta S},~~~~~
\dot{S} = \left\{ S,\tilde{H}\right\} =-\frac{\delta
\tilde{H}}{\delta P},
\end{equation}
where $\tilde{H}[P,S]$ is the {\it ensemble Hamiltonian} that generates time translations.

The fundamental variables of our phase space are the probabilities $P$ and the auxiliary function $S$. One may introduce the notion of an {\it observable\/} on this phase space, as any functional $A[P,S]$ that satisfies certain requirements \cite{HR2005, H2008}. For example, the infinitesimal canonical transformation generated by any observable $A$ must preserve the normalization of $P$. This implies the condition $A[P,S+c] = A[P,S]$; i.e., gauge invariance under $S\rightarrow S + c$.

Up to now, the discussion has been very general: since the ensemble Hamiltonian has not been specified, the formalism may be applied to a large class of statistical theories. The following ensemble Hamiltonians are of interest in that they lead to equations that describe the evolution of quantum and classical non-relativistic systems \cite{HR2005}:
\begin{eqnarray}\label{HaClQu}
\tilde{H}_C[P,S] &=& \int dx\, P \left[ \frac{|\nabla S|^2}{2m} + V(x)\right] ,\\ \nonumber
{~} \\
\tilde{H}_Q[P,S] &=& \tilde{H}_C[P,S]
+  \frac{\hbar^2}{4} \int dx\ P\frac{|\nabla \log P|^2}{2m} .
\end{eqnarray}
For example, the equations of motion derived from $\tilde{H}_Q[P,S]$ are given by
\begin{equation}\label{EqMoHaClQu}
\frac{\partial P}{\partial t} + \nabla .\left( P\frac{\nabla S}{m} \right) =0,~~~~~~~\frac{\partial S}{\partial t} + \frac{|\nabla S|^2}{2m} + V +  \frac{\hbar^2}{2m}\frac{\nabla^2 P^{1/2}}{P^{1/2}} = 0
\end{equation}
while the equations of motion derived from $\tilde{H}_C[P,S]$ are the same as Eq. (\ref{EqMoHaClQu}) but with $\hbar=0$. The first equation in Eq. (\ref{EqMoHaClQu}) is a continuity equation, the second equation is the classical Hamilton-Jacobi equation when $\hbar = 0$ and a modified Hamilton-Jacobi equation when $\hbar \neq 0$. Defining $\Psi:=\sqrt{P}~e^{iS/\hbar}$, Eq. (\ref{EqMoHaClQu}) takes the form
\begin{equation}\nonumber
i\hbar \frac{\partial \Psi}{\partial t}
= \frac{-\hbar^2}{2m}\nabla^2\Psi + V\Psi,
\end{equation}
which is the usual form of the Schr\"{o}dinger equation. Thus, in the ECS approach,
classical physics is not given a probabilistic description
in a phase space with coordinates $x$ and momenta $p$ using a phase space probability
$\rho(x,p)$. The two probabilistic descriptions are {\it not} equivalent in that in
general a $\rho(x,p)$ has to be described by a mixture of configuration space states
$P(x)$ and $S(x)$ \cite{L1952, RH2009}.

Notice that quantum and classical particles are treated on an equal footing in this more general framework, with differences being primarily due to the different forms of the respective ensemble Hamiltonians. One may ask whether the functions $P$ and $S$ can be interpreted in a similar way regardless of whether we are discussing a classical or a quantum system. We will show that this is indeed possible provided we do not try to assign properties to $P$ and $S$ that go beyond what is required of a {\it statistical theory}; i.e., these are quantities that should be used to describe the state of ensembles, to enable us to make predictions that can be compared to experiments.

Before looking at the role of probability, we consider the interpretation of $S$, which was introduced above as an auxiliary variable conjugate to $P$. One may define local energy and momentum densities in terms of $S$. If $\tilde{H}[\lambda P, S] = \lambda \tilde{H}[P,S]$, which holds true for the ensemble Hamiltonians of Eq. (\ref{HaClQu}), one can show that $\partial S/\partial t$ is a local energy density. Furthermore, $\int dx\, P\nabla S$ is the canonical infinitesimal generator of translations and therefore $P\nabla S$ can be considered a local momentum density. These results are {\it generally valid} \cite{RH2009}; i.e. they hold true for both classical and quantum systems and thus provide a common physical interpretation of $S$ that is appropriate for a statistical theory.

To maintain full generality, $S$ should not be regarded as a ``momentum potential'' for individual particles. In particular, for an ensemble of classical particles it is not necessary to assume that the momentum of a member of the ensemble is a well-defined quantity proportional to the gradient of $S$, as it is done in the usual deterministic interpretation of the Hamilton--Jacobi equation. Such an assumption would go beyond the requirements of a statistical theory and it is unnecessary. (This avoids forcing a similar deterministic interpretation in the quantum case which would correspond to the de Broglie-Bohm formalism. A deterministic picture can be recovered for classical ensembles precisely in those cases in which trajectories are operationally defined \cite{HR2005}.)

We have assumed that $P$ is a probability density; i.e., that it is possible to measure the state of the system and that $P(x') dV$ describes the probability of finding the system in the particular configuration $x'$ within the configuration space volume $dV$.

It is important to point out that such an interpretation of $P$ is generally valid (i.e., regardless of whether we are considering a classical or quantum system), despite the fact that it has been claimed that probability theory does not apply to quantum mechanics. Indeed, as B. O. Koopman pointed out in a seminal paper written in 1957, ``Ever since the advent of modern quantum mechanics in the late 1920's, the idea has been prevalent that the classical laws of probability cease, in some sense, to be valid in the new theory'' \cite{K1957}.

In his paper, Koopman goes on to refute this claim. In the introduction, he writes:
\begin{quote}
``The primary object of this presentation is to show that the thesis in question is entirely without validity and is the product of a confused view of the laws of probability. The situation can be straightened out at a very elementary level: all that is needed is to make quite clear that and explicit the concept of {\it event}. It will not be necessary to either to adopt any particular position regarding the controversial matters at the foundations of probability or to commit oneself at all deeply on the level of physical law.''
\end{quote}
Koopman's main point is that the claim that probability theory ceases to be valid in quantum mechanics is to a large extent the result of not distinguishing between {\it compatible} and {\it incompatible} events, but once this distinction is made, it can be seen that the axioms of probability theory are not violated in quantum mechanics. Events are interpreted in an operational sense, as is clear from the following quote:
\begin{quote}
``A thorough examination of all the concrete applications of the theory of probability shows that the concept of {\it event} can always be interpreted as {\it a statement concerning the state of a material system on a specified occasion}. It is essential, furthermore, that the statement be meaningful according to the simplest interpretation of Bridgman's operational standards: in principle, capable of verification (true or false) by an observation.''
\end{quote}
Koopman's article, which focuses on the double-slit experiment, seems to have been motivated by a publication of Feynman \cite{F1951}. The observations of Koopman do not seem to be very well known; most textbooks of quantum mechanics do not incorporate them (two notable exceptions are the textbooks of Ballentine \cite{B1998} and Peres \cite{P1995}).

An explicit proof that quantum mechanics satisfies the axioms of probability theory was given later by Ballentine \cite {B1986}, who used the standard representation of observables in terms of operators and verified, for both pure states and density matrices, that the axioms are satisfied. More recently, a similar result was obtained by Goyal and Knuth, this time using a different approach which allowed them to prove that Feynman's rules are compatible with probability theory by explicitly deriving Feynman's rules on the assumption that probability theory is generally valid \cite {GK2011}.

The misconception that quantum mechanics and probability theory are incompatible is unfortunately widespread. Ballentine reviews some erroneous applications of probability to quantum mechanics which have resulted in claims of inconsistency between probability theory and quantum theory. These typically involve mistakes where conditional probabilities are handled incorrectly. Probabilistic formulas that involve joint probability distributions also require some care because, as is well known, quantum mechanics lacks an expression for the joint probability distribution of variables whose operators do not commute. A formula that involves a joint probability distribution is no longer applicable if the joint probability distribution is not defined.

It is important to stress that the arguments discussed here are {\it independent} of  the choice of interpretation of probability, as Koopman already pointed out in his paper, because they are based on the axioms of probability theory and these are common to all interpretations. Therefore, the $P$ that is used in the description of statistical systems by means of ensembles on configuration space plays the {\it same} operational role independent of whether the ensemble describes a classical or a quantum system.

\section{A Statistical Description of Classical Physics}\label{stat}

One of the assertions of this paper is that classical physics is, in a very real sense,
a statistical model of nature.  It is, therefore, not the statistical character of
quantum mechanics that distinguishes it from classical physics but rather must be something else.
The ECS formalism presented in the previous section can be used to illustrate this
and provides as well a model for treating interacting quantum and classical systems
(see the Appendix).
This might tempt one to offer the ECS theory as a resolution to the ``classicality of experiments''
aspect of the measurement problem as well as providing a mathematical model for the Heisenberg cut.
However, we have already argued in Section \ref{SPME} that experiments
are described neither by classical nor quantum physics but rather by operational prescriptions that lie
outside both the formalisms of classical and quantum theory.  In fact, we will argue in Section \ref{MPCP} that, rather
than resolving any aspect of the quantum measurement problem, the statistical description of classical
physics brings to light a concomitant measurement problem in classical physics.

As we pointed out in the previous section, a statistical description of classical mechanics may be formulated using ensembles on configuration space. For example, in the ECS approach, the motion of a classical ensemble of particles under the influence of a potential $V(x)$ is described in terms of the ensemble Hamiltonian $\tilde{H_C}$ of Eq. (\ref{HaClQu}) and the resulting equations of motions are Eqs. (\ref{EqMoHaClQu}) with $\hbar=0$,
\begin{equation}\label{EqMotion}
\frac{\partial P}{\partial t} = -\nabla .\left( P\frac{\nabla S}{m} \right),~~~~~~~
\frac{\partial S}{\partial t} = - \frac{|\nabla S|^2}{2m} - V.
\end{equation}
In the limit of an initial $\delta $ function probability distribution, the
equations reduce to the exact classical equation of motion of a
single particle subject to no uncertainty. Therefore, this formalism
includes both the equation of motion of
an ensemble of particles and the exact equation of motion of a single
particle. In this sense, the above formulation might well be considered to be more
fundamental than the classical Newton's equations of motion.
The obvious interpretation is that $P(x,t)$ and
$[P(x,t)\nabla_x S(x,t)]$ describe the statistical distribution of the
results of measurements performed on an ensemble of similarly
prepared individual particles, which is analogous to the ensemble
interpretation of quantum mechanics.

At this point, it is important to stress that this statistical
description of classical mechanics, while unfamiliar to many, is not
particularly new or revolutionary. It is well known that the
Hamilton-Jacobi equation provides a formulation of classical
mechanics that may be considered as fundamental as the Hamiltonian
and Lagrangian formulations. It is also well known that given a
solution $S(x,t;c)$ of the Hamilton-Jacobi equation (where the $c$
are constants that specify the particular solution $S$), it is
always possible to associate with this solution a whole family of
conserved densities which satisfy the continuity equation.
Hamilton-Jacobi theory is fundamentally a theory of ensembles
\cite{L1952,S1961}.  Any normalizable density that
satisfies the corresponding continuity equation may be used to
describe a physically allowed classical state.

How is it that the probability density $P$ can be interpreted as a
fundamental description of a classical system? To be sure, all
measurements of systems, whether classical or quantum, are subject
to uncertainty but, for classical systems, these uncertainties are usually attributed to
{\it noise} in the experiment.  While, in principle, such uncertainties
can be made arbitrarily small, in practice this is not the case as
is well known to experimentalists. Rarely do the
experimental uncertainties associated with a measurement apparatus
even approach the fundamental limitations imposed by quantum
mechanics, e.g., $\delta p \delta x \geq \hbar$. In reality, the
{exact} physical states in classical physics are just
as inaccessible as they are in quantum mechanics. Nevertheless,
classical experimental uncertainties are routinely expressed as
``errors'' associated with the result of a measurement and rarely,
if ever, considered to be attributes of the system under
investigation and/or the measuring apparatus. But does this have to
be the case? If one considers the statistical formulation to be a legitimate
description of classical physics, then it appears that the
uncertainties associated with classical systems
should be as much a part of their theoretical descriptions
as are the fundamental uncertainties associated with quantum
mechanics.

One might ask why the uncertainties in classical physics are largely
attributed to the preparation and measurement processes rather than
to the theory. Perhaps, this happenstance is an historical accident.
The uncertainties might just as well been attributed to classical
theory as discussed above. However, in most cases the relative
uncertainties in classical systems are so small that ignoring them
or considering them to be uncertainties related to making
measurements is entirely understandable. If classical physics had
been generally concerned with very small (mesoscopic) systems or if
our present environment had been one that included a great deal of
randomly fluctuating forces, then this might not have been the case.
Even so, there are cases in which uncertainties are considered to be
part of the theory. One of the earliest such classical theories is
that of Brownian motion, but there are certainly other situations
where this arises. For example, predictions of the large-scale
structure in the universe are invariably statistical in nature due
to an inherent randomness in the state of the early universe. The
usual assertion is that the source of the randomness is quantum
fluctuations; however, this claim is more a matter of conjecture
than deduction.

There are aspects of the exact state formalism of classical mechanics
that are as troublesome as those of quantum mechanics.  Among the primary
theoretical constructs of classical physics are point particles, particle
trajectories, continuous media, and rigid bodies.  While these may be useful
approximations to what we observe in the physical world, their primary function
in classical mechanics is as part of the formalism that is used to predict
the statistical outcomes of experiments/observations.  Because the constructs
of classical physics were created from our everyday observations of the physical
world, it is understandable that we often identify them with reality whereas
in actuality they are simply part of the theoretical formalism that is necessary
for a self-consistent classical mechanics.

As soon as one moves from formalism to the physical world, it becomes
abundantly clear the theoretical constructs are fundamental different from
objects in the real world.  For example, the finite size, density, stiffness,
and viscosity of real bodies are all determined experimentally with no
fundamental {classical} understanding of how they arise. These properties
of matter are generally agreed to be quantum mechanical in origin
even though most are much too complex to be computed within quantum mechanics
and so are still determined experimentally \cite{fdr}. In fact, many aspects
of the consistency of classical physics collapse under close scrutiny because
of the underlying quantum nature of the phenomena. In any case, there seems
to be no compelling reason to cling to the precise, deterministic formulation
of classical mechanics in lieu of the statistical formulation that, in any
case, formally includes exact determinism by employing $\delta $ function
probability densities as pointed out above.

Perhaps a more problematic aspect of the statistical formulation is
the lack of a precise method of specifying the initial configuration
space probabilities of a system. On the other hand, recall that
neither is there a precise method for determining the quantum
mechanical state of a system without employing either the notion of
{preparation} or {measurement} of a system. As
discussed above in Section \ref{SPME}, quantum mechanical state preparation
usually proceeds via a set of operational prescriptions that have
been complied through the process of calibration. It seems
reasonable that the same language could be used to assign a
statistical distribution to a classical state. Certainly,
experimentalists spend a great deal, probably most, of their effort
in understanding and estimating uncertainties associated with the
experimental setup. These efforts might just as well be described as
determining the probability density of various components of the
system.

\section{The Measurement Problem in Classical Physics}\label{MPCP}

A statistical formulation of classical physics brings with it a
{classical measurement problem}. The probability function $P(x,t)$
only specifies the statistical outcome of an
ensemble of similarly prepared states. There is nothing in the
formalism to indicate that only one of the possibilities actually
occurs. Consider a position measurement: If one insists that the
probability function provides a description of an individual system,
then the original probability density must be updated (i.e., must
``collapse'')  to one that has
a lower uncertainty by some process associated with the measurement
(in general, the canonically conjugate function $S(x,t)$ will also require
updating). The description of this process must occur outside the
theoretical formalism of classical mechanics. Otherwise, as in the analogous
quantum case, the theory would predict the collapse and an exact
prediction could be made, negating the statistical nature of the
theory. So it seems that a statistical classical theory shares this
feature of quantum theory. Because
the contention is that physical theories, classical and quantum, are
statistical in nature, any implied ``collapse'' will necessitate
a process that is not included in the theory. It should be noted that, in the classical
case, there is no inherent linearity in the theory and so the
violation of linearity by the collapse is not the culprit.

It should be pointed out that this argument is not predicated on the
ECS formalism introduced in Section IV which is only an example, albeit a
compelling one, of how a statistical classical theory might be formulated.
The ``collapse'' problem would arise in any statistical theory of classical
physics.

Another aspect of the measurement problem that the statistical
classical theory shares with quantum mechanics is the Copenhagen
interpretation's requirement of the {classicality of
experiments}. At first glance, this may seem absurd; after all, how
could classicality be a problem for classical physics? However, what
is meant by the term {classicality} is not entirely clear.
For example, the following is one of Bohr's explanations of what he
meant by ``classical concepts'' (although it should be noted that
Bohr addressed this topic in many different ways, not all of which
were completely consistent) \cite{B1963}:
\begin{quote}
``The decisive point is to recognize that the description of the
experimental arrangement and the recordings of observations must be
given in plain language, suitably refined by the usual terminology.
This is a simple logical demand, since by the word `experiment' we
can only mean a procedure regarding which we are able to communicate
to others what we have done and what we have learnt.''
\end{quote}
Stapp chooses to emphasize this pragmatic view of classicality by
using the word {specifications}, i.e., \cite{S1972}
\begin{quote}
``Specifications are what architects and builders, and mechanics and
machinists, use to communicate to one another conditions on the
concrete social realities or actualities that bind their lives
together. It is hard to think of a theoretical concept that could
have a more objective meaning. Specifications are described in
technical jargon that is an extension of everyday language. This
language may incorporate concepts from classical physics. But this
fact in no way implies that these concepts are valid beyond the
realm in which they are used by technicians.''
\end{quote}
The point is that descriptions of experiments are invariably given
in terms of operational prescriptions or specifications that can be
communicated to the technicians, engineers, and the physics
community at large. In some sense, even the words we use to write
journal articles to present the results of an experiment might be
considered to be part of the measurement apparatus. Are these
operational prescriptions part and parcel of {classical
theory}? Are they couched in terms of point particles, rigid solid
bodies, and Newton's laws? Of course not. They are part of Bohr's
``procedure regarding which we are able to communicate to others
what we have done and what we have learnt.''

So why is it that the word ``classical'' can be taken in so
many ways by physicists? There seems to be a tendency among physicists
to think of every aspect of physics before quantum mechanics to be
part of the classical picture or, perhaps, every subject of any
scientific discipline that doesn't employ a quantum mechanical
analysis should be viewed as classical.  While in a certain sense
this is true, such a point of view tends to conflate the theoretical
language of classical mechanics with ordinary (albeit technical)
language, thereby removing the division between the theory of
classical physics and the description of physical measurements.
We have already pointed out that classical theory proper is quite
formal and contains theoretical constructs that are fundamentally
different from objects in the real world.  On the other hand,
most of ``classical physics'' is phenomenological and makes no
pretense to being fundamental. Classical physics does not
represent an all encompassing, although incorrect, theory of the
world.  It is more patchwork of theoretical classical mechanics,
electromagnetism, and phenomenological truths combined with the
conventional wisdom and standard prescriptions of experimental
physics.

So how do we treat measurements in classical physics? The same way
as we do in quantum mechanics, with operational prescriptions in
plain language so that the results can be communicated to the
scientific community. These prescriptions are not contained within
classical theory. In this sense, the description of a measurement in
classical physics must be in terms of language that falls outside
the theory. Of course, the measuring apparatus itself can be
described (statistically) in terms of the (sometimes
phenomenological) concepts of classical physics in the same way that
it can be described quantum mechanically (although it rarely
is). But to the extent that the apparatus is treated as part of the
classical (quantum mechanical) system, it becomes part of the
system under investigation and can no longer be considered part of
the measurement. Heisenberg expressed this in the extreme
(quoted in Ref. \cite{SC2008}): ``One may treat the whole
world as one mechanical system, but then only a mathematical problem
remains while access to observation is closed off.''

The bottom line is that classical physics is faced with the same two
major aspects of the measurement problem as quantum mechanics: 1)
The theory is fundamentally statistical in nature and any attempt to
interpret it for single systems requires a ``collapse'' that
necessarily lies outside the
theory; and 2) The descriptions (specifications) of experiments must
be operational prescriptions that are outside of theoretical
physics, and the results must be communicated in plain language
rather than with theoretical concepts. Our pedestrian approach to
the former is simply that both quantum and classical physics are
theories, not of individual systems, but rather of the statistical
behavior of ensembles of systems. The quantum part of this statement
is consistent with the minimalist interpretation of quantum
mechanics espoused by Einstein \cite{E1949}, Ballentine \cite{B1970}, and
others. As for the latter
aspect of the measurement problem, our pedestrian approach is
similar to Bohr's and Heisenberg's as espoused in the Copenhagen
interpretation: ``the description of the experimental arrangement
and the recordings of observations must be given in plain language''
and not in terms of theoretical constructs.  (However, it should be
noted that Bohr used the terms `in plain language' and `classical
physical concepts' interchangeably.)  Without these operational
descriptions, both quantum mechanics and classical theory are
mathematical formalisms that necessarily remain detached from the
real world.

The description of the classical measurement problem presented
here leads necessarily to a reevaluation of some of the issues raised
in Section \ref{INTRO} regarding the quantum measurement problem.
Part of the classicality of experiments aspect of the quantum measurement
problem is the issue of where to place the quantum/classical divide.
We pointed out above, locating the divide goes only part way. One must
also introduce a method to bridge the two formalisms, quantum and classical,
in order to describe the occurrence of a measurement.  Decoherence theory
avoids this aspect of the measurement problem by treating the measuring
apparatus as another quantum mechanical system. However, in so doing,
the actual measurement is simply pushed out further until it can be
described by operational prescriptions that define it. The divide between
the theoretical predictions and the measurement remains unaccounted
for. The ECS formalism (see the Appendix) is capable of dealing
with interactions between one system described by quantum formalism
and another by classical formalism; however, here again, the description
of the operationally defined measurement falls outside both quantum
and classical theory. This topic is addressed in Section \ref{Ope}.

Another aspect of the measurement problem listed in the
introduction, ``the consistency of quantum and classical reality''
must be rephrased as ``the consistency of quantum and classical
physics with reality.'' That is, how do we reconcile the statistical
nature of quantum and classical physics with our observed
perceptions of the world. Here again the minimalist (ensemble)
interpretation of our pedestrian approach obviates this problem by
restricting the predictions of both classical and quantum physics to
the statistical outcomes of measurements performed on ensembles of
similarly prepared systems. The modesty of this restricted
interpretation of the domain of physical theory will certainly be
distasteful to many physicists. We will comment on this in Section \ref{D}.

\section{Operational Prescriptions, Classicality, and Sufficient Precision} \label{Ope}

If the above claim of the statistical nature of both quantum and
classical physics is taken seriously, then perhaps we are in need of
a more general Copenhagen-type interpretation. A possible version of
such was alluded to in the previous section; however, questions
remain. For example, what is it that defines the transition from the
theoretical evolution of the system to the measurement, which is
defined operationally? For Bohr, Heisenberg, and Pauli, the
demarcation is simply when one stops talking about quantum mechanics
and starts talking about classical physics, with the acknowledgement
that this transition is, to a certain extent, at the discretion of
the experimenter.

However, now quantum and classical interacting systems, taken
together, are considered to be part of the theoretical (statistical)
evolution of the combined system. Then what constitutes a
measurement? A crucial aspect of a measurement is that it
be the declaration of a {precise} result \cite{precision}, which can only be compared
to the statistical predictions of the theory.  The ensemble
interpretation characterizes these predictions as corresponding to
the distribution of measurements made on an ensemble of similarly
prepared systems provided multiple measurements are possible; when this can not be realized, it is still possible to interpret the prediction in a
Bayesian sense, i.e., where probability is defined using the notion
of degree of belief.  How do we normally interpret experimental
uncertainty? In general, a one standard
deviation (1$\sigma$) value, for example, reflects the belief that
were one to repeat the experiment many times, the results would be
scattered about a mean value with $\sim 68 \%$ of the results
falling within 1$\sigma $ of the mean. This is precisely the
prediction made by the statistical ensemble interpretation of a
classical experiment. In some cases, experiments are actually
performed many times and the quoted uncertainties represent the
statistical distributions of the {results}. In this case, it
is usually the {standard deviation of the mean} that is
quoted as the measurement uncertainty, which is interpreted as the
spread in the probability distribution of the mean of the results of
multiple measurements if these were to be repeated in an ensemble
of multiple measurements.  In the case that an experiment cannot be
repeated, e.g., the determination of the large-scale structure of the
universe, then one must interpret predictions as degree of belief.

In cases where uncertainty is established by multiple measurements,
one might wonder how it is that the statistical properties of the
{results} of multiple measurements can be somehow attributed
to the statistical state of the system instead of interpreted as
uncertainties in the measurement. However, this situation can simply
be regarded as a {calibration} that enables the specification
of the statistical state of the system. The same circumstances can
occur in a quantum mechanical system that is determined, after the
fact, by multiple experiments on similarly prepared systems. In
other cases, when the predominant uncertainty can be
predicted or measured before hand, then one can simply assign a
probability distribution to the system. Of course, there may be many
sources of uncertainty that require distinct probability functions to be
assigned to different parts of the system, which are then allowed to
interact with one another. Recall that components of the
``measuring'' apparatus are to be considered as components of the
whole system under investigation.

If the ``measuring apparatus'' is to be considered part of the
combined quantum/classical system, what is the actual
{measurement} that is assumed to lie outside of the
theoretical description of the system, i.e., what marks the division
between the theoretical description of the system and the
operational description of the measurement? The answer to this
question brings us to the notion of {\it precision}
\cite{precision}. The contention is that the measurement occurs at
the point in the evolution of the system at which there is no
further uncertainty that may affect
repeatability. Then the result is of {sufficient precision}
that a measurement has been made. This is an imprecisely defined
transition in that the specification of sufficient precision is left
up to the experimenter. If the scientific question being addressed
requires more precision then a part of the sequence that was
formally considered an operational description of the measurement
might, instead, have to be specified probabilistically and be
included in a theoretical treatment of the system. In some ways,
this is analogous to the imprecisely defined quantum/classical
divide of the Copenhagen interpretation.

Once a {precise} measurement has been made, it has a
permanence that can then be communicated to others visually, in
writing, verbally, in technical language, in English, in Spanish,
etc. Because measurements are considered to be exact and permanent,
they become a matter of record. However, even at this point one
might wish to introduce uncertainty into the process. What if one
wishes to account for transcription errors, errors in graphical
illustration, errors in language translation, linguistic errors,
interpretational errors, erratic human behavior, etc.? In principle,
these aspects of communication might also be treated theoretically
using information theory, with specific probability distributions
determined by calibration and then such communications could be
considered to be part of the theoretical evolution of the system.
However, in this case, it is doubtful that one would characterize
this part of the system as belonging to classical theory.  At some
point one simply draws the line after which errors are dismissed with
the proverbial ``mistakes were made.''

So far, we have discussed the evolution of a quantum measurement as
the following sequence: 1) follow an operational prescription to
prepare the quantum mechanical state of the system and the classical
state of the measuring apparatus; 2) determine the quantum
mechanical evolution of the system; 3) determine the joint
quantum/classical evolution of the system and measurement apparatus
(if an ECS type formalism is available) or the joint quantum evolution
of the system and measuring apparatus (if a decoherence analysis is
performed); 4) determine the classical evolution of the apparatus; 5) follow
operational prescriptions to determine the {precise} result of
the measurement; and 6) communicate the result to others. Of course,
measurements needn't proceed in such a linear fashion. There can be
many quantum and classical components of the system and these can
interact at different times resulting in complicated mixed
quantum/classical states. Or, it might be possible to prepare a
quantum state, let the state evolve, and then perform the
``measurement'' following a set of operational prescriptions without
{\it any} involvement of a truly classical system (as decoherence theory
would maintain); however, no specific examples of such experiments come
immediately to mind.  With regard to 3) above, even in the absence of
a consistent model of quantum/classical interactions, it is undoubtedly the
case that approximate (semi-classical) models are available with sufficient accuracy to
represent the evolution of the systems.

One topic we have not touched upon is the relation of quantum mechanics and information.
As far back as Bohr and Heisenberg, there have been interpretations of quantum mechanics
that emphasized the relation of the quantum wave function to an experimenter's knowledge
of or information about the real world as opposed to postulating that the wave function
is a representation of reality. The operational prescriptions discussed above should then
be described as prescriptions for increasing an observer's knowledge about a given system.
The issue of how the operational prescriptions are related to an increase in our knowledge
is important, but it is outside of the scope of this paper. At a minimum, it would require
bringing in concepts from information theory (to quantify the amount of information provided
by an experiment which is carried out according to an operational prescription) and probability
theory (to quantify how the uncertainty is diminished by the data). These fields provide tools
that are extremely important for the description of the experimental issues that play such a
fundamental role in both classical and quantum measurement theory. While some important aspects
of the operational prescriptions will very likely remain recipes that cannot be described in a
mathematical way, efforts to provide a clearer description of those aspects that can be formulated
in the more formal language of information theory and probability theory are needed. However, we believe that such attempts at rigorous formulations should not be carried out in such a way
that the complex experimental issues encoded in operational prescriptions are ignored.

\section{Discussion}\label{D}

As alluded to earlier, it is curious that the measurement problem
has persisted over eight decades even while advances in fundamental
physics, quantum field theory in particular, have been remarkable to
say the least. Perhaps one reason for its persistence is because
physicists have sought theoretical solutions to what is essentially
an experimental problem. As a consequence, in debates about the measurement problem,
the experimental side of physics is largely ignored except in
sweepingly general statements about experiments that, if not
inaccurate, are certainly of limited validity. Many quantum
interpreters seem to believe that fundamental physical theories
should provide an accurate (or as an accurate as is possible) representation of reality and that
the theories themselves should provide the basis for interpreting
experiments. Many experimentalists would not necessarily subscribe
to this principle. In this respect, the following quote by
Gutzwiller \cite{G1990}
seems particularly relevant:
\begin{quote}
``The discussion of the so-called `thought experiments' in most
cases is singularly crude, i.e., removed from any awareness of the
practical considerations in a real experiment. Time and effort is
spent on purely mathematical relations. With few exceptions, the
hard work of writing down, and then solving the relevant equations
for a specific laboratory set-up has not even begun; in particular,
the inevitable presence of noise is ignored most of the time.''
\end{quote}

In the early days of quantum mechanics, its domain was essentially
limited to the microscopic world, i.e., the world of atoms. The
Copenhagen interpretation was often expressed in statements like,
``quantum mechanics is a theory that makes statistical predictions
about measurements made on microscopic systems with classical,
macroscopic measuring apparatus.'' It seems that Bohr, Heisenberg,
and Pauli were comfortable with this situation and were willing to
limit the applicability of quantum mechanics to microscopic (atomic)
systems. Nevertheless, there are significant reasons to consider
quantum mechanics to be much more fundamental than classical
physics. Today, most physicists view the quantum theory of matter
and radiation as a truly {fundamental} theory of nature.
About this, there can be little doubt. Because quantum coherence
is critical to the measurement problem, physicists often point to
macroscopic coherent quantum behavior, e.g., superconductivity and
superfluidity, to demonstrate that quantum mechanics applies even to
large scale structures. One needn't, however, appeal to such exotic
phenomena. In a very real sense, the phase information in
electromagnetic waves is due to quantum coherence in a Bose
condensation of photons. Moreover, virtually every property of
ordinary macroscopic matter can be understood only in the context of
quantum theory. On the other hand, in classical physics nearly every
property of matter is understood only phenomenologically. Indeed,
the behavior of an ordinary metal spring that obeys the
``classical'' Hooke's law can also be taken as observational
evidence in support of quantum mechanics because without quantum
mechanics the very quality of the stiffness of the spring is an
unexplained phenomenon.  There are certainly less fundamental 
quantum theories (e.g.,
the non-relativistic Schr\"{o}dinger equation) that follow from
more fundamental theories (the relativistic Dirac equation); 
however, for example, the classical Hooke's law doesn't follow from
any classical theory.
To be sure, both classical electrodynamics and general relativity
were considered to be classical fundamental theories; however,
the former of these is known to be only the classical limit of
the more fundamental quantum electrodynamics.  The same is usually
assumed to be true for general relativity even though as yet there
is no accepted quantum theory of gravity.

These arguments together with the great predictive successes of
quantum mechanics and quantum field theory (as embodied by the
twelve decimal place agreement of the quantum electrodynamical
prediction with the measured g-2 value of the electron) have
probably led most physicists to accept that everything in nature is
quantum mechanical in origin and to infer that classical physics
corresponds to an approximation to quantum theory that is quite
accurate in the limit of macroscopic systems with large energies. In
fact, nothing that has been discussed in the present paper would
contradict this view. This isn't to say that classical theory isn't
enormously useful in our description of the world. In fact, the
majority of calculations in physics are surely classical
computations. Furthermore, there is little doubt that quantum
mechanics would be absolutely useless in effecting solutions to
these problems; the complexity of most phenomena assures us of this.
So how do we answer the Einstein-type question about the nature of
the real world? Does the fundamentalness of quantum mechanics trump
the usefulness of classical physics or should we be content with
Bohr's contention that both quantum (microscopic) and classical
(macroscopic) accounts must be employed?

At this point, it may be prudent to remind ourselves of what we
often tell our students and the public for that matter. That is, our
theories should be considered to be simply ``models'' that are
useful devices in describing the real world. If one were to ask most
scientists whether a particular model coincided with the ``real
world,'' the answer would probably be ``no, they are just models''
\cite{conventionalism}. It is the fact that physical theories
(models) are amenable to change in the face of new evidence that, in
large part, is the strength of the discipline. Granted, some models
are extremely good. But as good as they are, they are only models
which we use to make sense out of the natural world and often have
only a limited domain. If this is the case, then why should we be
upset by the incongruous union of quantum theory and classical
experimentation? Here again, the 12 decimal place agreement between
quantum electrodynamical theory and experiment leads many to take
the position that quantum theory, or some future version of it, is
the embodiment of reality. However, one should be wary of
deeming a single (or even several) high precision measurement(s)
as strong evidence in support of a theory. \cite{support} On the other hand, if one
takes seriously the view that quantum theory is simply a model of
reality with a necessarily limited domain of applicability, then
perhaps the Copenhagen interpretation isn't so distasteful.

Some might object to referring to quantum mechanics as ``simply a model of reality'' with a ``limited domain of applicability.''  Because quantum mechanics deals with the fundamental building block of nature, is it not surely the fundamental theory from which all others follow?  The problem is that other theories don't necessarily follow from quantum mechanics nor is quantum mechanics at all useful in dealing with most of the phenomena in nature.  Consider, for example, the problem of predicting the motion of an irregularly shaped solid object tumbling in free space.  A solution to this problem is relatively straightforward in terms of classical mechanics; although, the addition of the elastic properties of the solid would complicate the problem significantly.  Of course, the elastic properties of the object are phenomenological at the classical level and can only be derived from first principles from the quantum mechanical behavior of the elementary particles that make up the solid.  On the other hand, solving for the motion of the solid from a field theoretic account of the behavior of these particles is essentially impossible.  In fact, it’s not even clear that quantum field theory is capable of formulating the problem. Instead, we invent a model for the behavior of macroscopic solid objects, classical mechanics, which provides extremely accurate predictions in such instances.  If one moves from such a simple system, an irregular solid body, to systems of extreme complication, e.g., human behavior, the connection to quantum mechanics is even more remote.  Finally, there is the fact that the predictions made by quantum mechanics (or classical mechanics for that matter) require interpretative statements and observational prescriptions that lie outside the theory.

One might argue, from a reductionist point of view, that surely the theory of the ``smallest'' components of matter and energy, out of which everything in the physical world is composed, occupies a privileged place in natural philosophy.  The fundamental principles inherent in physics in general and quantum mechanics in particular are surely to be thought of as fundamental principles of nature.  In a sense, this is true; however, in reality these principles are the rules governing the models we construct and not of nature herself.  For example, Newton’s universal law of gravitation, embodied by $F = Gm_1m_2/r^2$ and $F = ma$, are rules that govern the behavior of the constructs of the theory, massive point particles and the trajectories of those particles.  There are no point particles and exact trajectories in nature.  These are theoretical constructs of a model that has proved to provide an extremely useful description of some aspects of nature.  The same is true for quantum mechanics, the principles of which govern the deterministic behavior of wave functions (or Hilbert state vectors), the theoretical constructs of the theory.  This model has proved incredibly useful in describing the behavior of matter and radiation (when combined with interpretive statements and observational prescriptions), but the principles of quantum mechanics are not, in themselves, laws of nature but rather the rules that govern the theoretical constructs of a particular model of nature.

In fact, it is the mistaken identification of the theoretical constructs of a theory with entities in the natural world that sets the stage for the measurement problem.  Although we have argued that classical mechanics has its own measurement problem, it is understandable why this is usually overlooked.  The constructs of classical mechanics, e.g., point particles, solid bodies, and continuous media, all bear a strong resemblance with objects we see in nature and so we tend to identify these constructs with the reality that they are intended only to model.  In quantum mechanics, following this practice of identifying theoretical constructs with reality immediately leads to serious conceptual difficulties.  To what objective reality does a wave function or Hilbert space vector correspond?  If we identify wave functions with objective reality we are inexorably led to the measurement problem.  As to whether or not quantum mechanics is the embodiment of reality, the above arguments suggest that quantum mechanics is neither complete nor does it even provide an accurate description of reality. It is, rather, an extremely powerful and useful model which helps us understand the physical world around us. 

One of the most pernicious aspects of the measurement problem is
the imprecisely defined quantum/classical divide.  It is largely the
inability of the Copenhagen and von Neumann interpretations of quantum
mechanics to model the quantum/classical interaction that renders them
incapable of resolving the measurement problem.  In our pedestrian
approach, we argue that it is not the quantum/classical divide that
is problematic but rather the divide between the quantum/classical
analysis and the operational prescriptions that characterize the
measurement process.  The fact that this latter divide is not
mathematically describable is not, we believe, a problem because
the operational prescriptions themselves are not theoretical constructs
of physics but rather are given in Bohr's "...plain language...to
communicate to others what we have done and what we have learnt."
We do not claim that this distinction resolves the measurement problem
but rather that it reveals that the measurement problem is not a theoretical
problem to be resolved.  The realization is that all physical theories,
quantum and classical, derive their meaning by appealing to concepts that
must lie outside the theory.  On the other hand, our pedestrian
approach must still deal with the interactions between the quantum system and
the classical measuring apparatus that take place prior to the measurement.
The formalism discussed in the Appendix provides an alternative, self-consistent 
way to treat the interaction of a quantum system with a
classical measuring apparatus.  However, like decoherence theory, this
formalism is much too difficult to apply in all but the simplest
cases.  Therefore, one cannot claim that either of these two formalisms
could have provided much help for physicists in analyzing the results
of a given experiment. It seems reasonable that the liberal utilization of
pragmatic, semi-classical, and heuristic arguments in the context of understanding
complex systems has served to finesse such situations.

Because some physicists assert that decoherence theory has largely
resolved the measurement problem, it is important to point out the
limitations of this claim. As discussed above, decoherence theory
does demonstrate why it is that macroscopic measuring apparatus,
even if treated quantum mechanically, behave, statistically,
according to notions of classical physics. Whether one chooses to
perceive this as verifying how classical physics {emerges}
from quantum mechanics or only that classical behavior is
{consistent} with quantum mechanics, the primary dilemma of
the measurement problem remains. That is, how is it that a
continuous, evolving wave function leads to our immediate perception
that specific events occur in space and time. Finally, while
decoherence theory can explain how the
reduced (averaged over environmental degrees of freedom) density
matrix implies the same behavior as a proper mixed state, it is
silent on the meaning and origin of proper mixed states. Of course,
this is a problem of quantum mechanics in general, that is, mixed
states are never the result of quantum evolution but rather must be
posited initially using a combination of quantum and classical
arguments. In short, in the context of quantum theory, where does
ordinary experimental uncertainty arise? Finally, as we argued
above, even if these issues are dismissed, in the end we encounter
operational prescriptions expressed in plain language about which
decoherence theory has nothing to say.

A primary message of the present paper is that classical physics is
subject to the same measurement dilemma as quantum mechanics. The
resolution is, in short, that both quantum mechanics and classical
physics are probabilistic in nature and can only be interpreted as
providing statistical interpretations of the outcomes of
measurements made on ensembles of similarly prepared systems. In
addition, the experiments themselves are ultimately described
operationally in plain language and the outcome of any given
experiment is a {sufficiently precise} result that can only
be understood in the context of a predicted statistical
distribution. In this sense, our pedestrian approach is
very similar to that provided by the Copenhagen interpretation of
quantum mechanics. As to whether the evolution of the systems is
purely quantum mechanical, purely classical or a mixture of the two
is, to a degree, at the discretion of the observer. Where does this
leave us with respect to the criticism that
a theory which only describes the statistical behavior of an
ensemble of systems cannot be considered a complete description of
reality? One is left with two options:
Either the pedestrian approach has resolved or perhaps clarified
the measurement problem or the measurement problem is even more pervasive
than before, encompassing classical as well as quantum physics. We choose
to embrace the former.

Suppose one were to accept the above pedestrian
approach as a solution to the measurement problem. This would
certainly imply that quantum mechanics cannot provide a complete
description of reality nor can a combination of quantum and
classical physics. Wouldn't this constitute a crisis, the result of
which might be a serious impediment for theoretical physics? This is
doubtful. As pointed out above, the quantum measurement problem has
been with us for the last 80 years and yet it seems not, in the
least, to have impeded the advance of quantum physics. Physical
theories are models we construct to comprehend nature. The fact that
they are limited in their validity should not necessarily be
considered problematic. That some such theories rely on ``classical
physics'' or the art of experimentation should not be taken as
detracting from their validity.  On the other hand, neither do we
suggest that searching for new ``resolutions'' to the measurement problems
is necessarily a futile endeavor.
Certainly past efforts have contributed to a deeper understanding of
fundamental physics and we undoubtedly have not learned all there is
to know about quantum mechanics.  Also, it's conceivable that some future
theory, some extension of quantum mechanics, will be capable of explaining
everything in terms of the fundamental concepts of the theory; however,
we doubt that this will ever happen.

Quantum electrodynamics (and its 12 decimal place accurate
prediction of the g-2 factor of the electron) is a fantastic
confirmation of the success of a fundamental physical theory
regardless of whether this theory relies on other products of human
imagination for its confirmation. We admit that one of the
incentives for our current views on the meaning of quantum mechanics
has been a desire to present gravity as a completely classical
phenomenon with no quantum aspects. However, even if superstring
theory turns out to provide a unified quantum view of all the four
forces of nature, including gravity, it will not affect the
arguments given above. Such a unified theory of everything would,
undoubtedly, still be framed in terms of probabilities and would
still depend on the operational description of experiments to derive
its meaning. Does this qualification make such a ``theory of
everything'' less prized? No. If such a theory is ever constructed
it would still rank among the greatest achievements of mankind.

We pointed out in the abstract that past attempts to resolve the
measurement problem have led to significant advances in our understanding
of quantum mechanics with important consequences in the mesoscopic
domain.  If the measurement problem were to be categorized as a faux
problem as this paper suggests, would that not have the effect of stifling
investigations of the interactions of microscopic, mesoscopic, and
macroscopic systems?  This is highly doubtful.  The current advances in the
understanding of such interactions and their applications to quantum
computing, molecular biology, nanophysics, etc., have been extraordinary.
These successes have spawned a vigorous, ongoing enterprise whose
proponents rarely, if ever, frame questions in terms of the measurement
problem.  In fact, reframing the measurement problem as we have in this paper
might well motivate new and fruitful investigations of the foundations of
quantum mechanics.

Physics is often said to be an experimental (and observational)
science. The real world confronts us every day with a
plethora of phenomena and it is the job of science, physics in
particular, to construct models to help us to understand what we
observe as well as to predict the outcome of various situations.
Because of the vast extent of nature, this understanding quite
naturally takes the form of a {patchwork of complementary
models} \cite{S1972}. The quest for a single, all encompassing,
consistent theory that provides a complete and accurate description
of the entirety of nature strikes us as the extreme of hubris. It's
not necessarily useful or even desirable to view nature as evolving according to
a single fundamental law. What we can do is to
construct improved models to make sense of the universe. That we
have been so successful in doing so is already nothing short of
amazing, as is expressed in the well-known declaration
(often attributed to Einstein), ``The
most incomprehensible thing about the world is that it is
comprehensible.'' This view of nature isn't held by all scientists
and certainly not by all physicists. (We refer you to M. Tegmark's
notion that the real world \textit{is precisely} its
mathematical structure \cite{T2008}.) The current patchwork of models leaves open
the possibility that future theories (models), quantum or classical,
may be capable of describing individual systems in the way that our
current models do not. However, at this time it seems that such
theories are not required and that, at least until future experiments
raise new possibilities, the current (patchwork) structure still has
a great deal to offer and that many new and exciting phenomena and
theories will be discovered within this context.

\appendix
\section{Coupling of classical and quantum systems}\label{ECS}

We have seen in Section \ref{PAQM} that both quantum and classical mechanics
may be described using the same theory of ensembles on configuration space.  
The formalism allows for the
coupling of classical and quantum systems in a natural and
self-consistent way \cite{HR2005, H2008, RH2009}. For example, consider the interaction of a
quantum particle with mass $m_{q}$ and configuration space
coordinates $q$ and a classical particle of mass $m_{x} $ and
configuration space coordinates $x$. Further suppose that the
interaction between the two particles is represented by a potential
$V(q,x,t)$ that depends on both classical and quantum coordinates.
Then the total  ensemble  Hamiltonian for the system becomes
\begin{equation}\label{HQC}
\tilde{H}_{QC}[P,S] = \int dq\,dx\, P\,\left[ \frac{|\nabla_x S|^2}{2m_x}
+ \frac{|\nabla_q S|^2}{2m_q} + \frac{\hbar^2}{4} \frac{|\nabla_q \log P|^2}{2m_q} + V  \right].
\end{equation}

There have been many suggestions on how to couple quantum and
classical systems as well as many arguments as
to why the coexistence of quantum and classical systems is
inconsistent, which are offered as ``proofs'' that the real world
must be entirely quantum mechanical in origin.
Most of the latter arguments have been shown to be fallacious or to
lack generality while specific models of
quantum/classical interactions, for the most part, have serious
shortcomings (see the discussions in \cite{HR2005} and \cite{H2008}
and references therein). The ECS
formalism overcomes these problems, e.g., it allows for back
reaction on the classical system, positivity of probability,
conservation of both probability and energy, the correct equations
of motion in the classical limit, the correct equations of motion
for both classical and quantum systems in the limit of no
interaction, automatic decoherence of quantum ensembles, the
uncertainty relations for conjugate quantum variables, and seems
capable of providing descriptions of physically interesting
interactions \cite{HR2005,RH2009,H2008}. In short, it
provides a consistent model for classical-quantum interactions.

In this paper we are interested in the ECS formalism to the
extent that it provides a model of measurement that may be used to
describe in a consistent way the interaction between a classical
apparatus and a quantum system that is being measured. The equations
of motion that follow from the ensemble Hamiltonian $\tilde{H}_{QC}$
of Eq. (\ref{HQC}) are non-linear. However, if the coupling between
the classical apparatus and the quantum system is weak, which would
be the case for a measurement that does not disturb the quantum system
too much, then the departure from linearity will be minimal for the
quantum system. Of course, non-linearity would be expected for any
formalism that deals with the measurement problem. After all, wave
function collapse is non-linear as, in some sense, is the Copenhagen
interpretation which imposes a statistical meaning for the wave
function through the non-linear Born rule, $P \sim |\Psi|^2$.
Because of its inherent non-linearity, the ECS formalism can be
extremely difficult to deal with computationally. Therefore, it
seems highly unlikely that it will ever serve as a general
computational tool with which to solve general quantum mechanical
{measurement} problems. Nevertheless, it does provide an
extremely useful analysis for foundational problems in physics. In
addition, because it treats quantum and classical phenomena on equal
footings, it could provide a vehicle for classical models of
fundamental physics forces, for example gravity \cite{B2009}.

As a successful model of classical-quantum interactions, the ECS
formalism does reduce some of the vagueness of the Copenhagen
interpretation and, therefore, offers some answers to the
measurement problem. There would no longer be any question as to the
location of the quantum-classical divide. It is located precisely
when and where the classical-quantum Hamiltonian specifies it to be
\cite{H2008}. Even so, Heisenberg's view that the divide is, in some
sense, at the discretion of the experimenter is still tenable
because of the inherent freedom in constructing a sensible
Hamiltonian relevant to a particular experimental setup. In other
words, different choices of classical-quantum coupling terms are
possible. The Schr\"{o}dinger cat paradox doesn't arise because the
cat belongs to the classical side of the system and is not,
therefore, represented by a wave function. Interference and the
superposition principle are not part of the classical world. The
classicality of experiments is no longer a problem because there is
now a completely classical configuration space in which the
measuring apparatus resides. In fact, it would seem that the ECS
model provides a complete solution to the measurement problem;
however, the very probabilistic description that allows classical
and quantum mechanical systems to be treated side by side brings
with it a {classical} measurement problem, on which we
elaborated in Section \ref{MPCP}.


\begin{thebibliography}{99}
\bibitem{precision} We use precision in its technical sense; i.e., reproducibility or repeatability, or the degree to which repeated measurements made under the same conditions are consistent with each other.
\end{thebibliography}
\end{document}